\documentclass[review,12pt]{elsarticle}

\usepackage{amssymb}
\usepackage{amsthm}
\usepackage{framed} 
\usepackage{amsmath,color}
\usepackage{mathrsfs}
\usepackage{graphicx}
\usepackage{epstopdf}
\usepackage{float}
\usepackage{caption}
\usepackage{subcaption}
\usepackage{bm}
\usepackage{bbm}
\usepackage{mathrsfs}
\usepackage{cleveref}
\usepackage{soul}
\usepackage{accents}
\usepackage{color,soul} 
\usepackage{color} 
\usepackage{nomencl} 
\makenomenclature 
\biboptions{sort&compress}
\soulregister\citep7 
\soulregister\citet7 
\soulregister\citealp7 
\newsavebox{\measurebox} 

\journal{Materials Science and Engineering A}

\makeatletter
\def\@author#1{\g@addto@macro\elsauthors{\normalsize%
    \def\baselinestretch{1}%
    \upshape\authorsep#1\unskip\textsuperscript{%
      \ifx\@fnmark\@empty\else\unskip\sep\@fnmark\let\sep=,\fi
      \ifx\@corref\@empty\else\unskip\sep\@corref\let\sep=,\fi
      }%
    \def\authorsep{\unskip,\space}%
    \global\let\@fnmark\@empty
    \global\let\@corref\@empty  
    \global\let\sep\@empty}%
    \@eadauthor={#1}
}
\makeatother

\begin{document}

\begin{frontmatter}



\title{Fracture toughness characterization through notched small punch test specimens}


\author{Emilio Mart\'{\i}nez-Pa\~neda\corref{cor1}\fnref{Uniovi}}
\ead{mail@empaneda.com}

\author{Tom\'{a}s E. Garc\'{\i}a\fnref{Uniovi}}

\author{Cristina Rodr\'{\i}guez\fnref{Uniovi}}

\address[Uniovi]{Department of Construction and Manufacturing Engineering, University of Oviedo, Gij\'on 33203, Spain}

\cortext[cor1]{Corresponding author. Tel: +34 985 18 19 67; fax: +34 985 18 24 33.}

\begin{abstract}
In this work a novel methodology for fracture toughness characterization by means of the small punch test (SPT) is presented. Notched specimens are employed and fracture resistance is assessed through a critical value of the notch mouth displacement $\delta^{\mathrm{SPT}}$. Finite element simulations and interrupted experiments are used to track the evolution of $\delta^{\mathrm{SPT}}$ as a function of the punch displacement. The onset of crack propagation is identified by means of a ductile damage model and the outcome compared to the crack tip opening displacement estimated from conventional tests at crack initiation. The proposed numerical-experimental scheme is examined with two different grades of CrMoV steel and the differences in material toughness captured. Limitations and uncertainties arising from the different damage phenomena observed in the lowest toughness material examined are thoroughly discussed. 
\end{abstract}


\end{frontmatter}



\begin{framed}
\nomenclature{$\delta$}{crack tip opening displacement}
\nomenclature{$\delta_c$}{crack tip opening displacement at crack initiation}
\nomenclature{$\delta_{IC}$}{crack tip opening displacement fracture toughness}
\nomenclature{$\delta^{\mathrm{SPT}}$}{notch mouth opening displacement of the small punch notched specimen}
\nomenclature{$\delta^{\mathrm{SPT}}_c$}{notch mouth opening displacement of the small punch notched specimen at crack initiation}
\nomenclature{SPT}{small punch test}

\printnomenclature
\end{framed}

\section{Introduction}
\label{Introduction}

The mechanical characterization of industrial components by means of conventional methodologies is, in many engineering applications, an extremely complicated - or even infeasible - task. These are, e.g., the cases of a structural element with complex geometry, small size (with respect to standard testing specimens) or that requires to be characterized without compromising its remaining in-service life. Furthermore, some particular applications require a continuous structural integrity assessment from a limited amount of material. This is the case of reactor pressure vessels, where the characterization of irradiated materials is hindered by the restricted number of specimens available. Hence, small scale techniques and micromechanical damage models have been developed with the aim of estimating mechanical and fracture properties while optimizing resources. From the modeling perspective, accurate toughness predictions have been obtained by extracting model parameters from Charpy V-notch and uniaxial tensile tests \cite{S94,SU94}. While on the experimental side, a significant progress has been achieved with the Small Punch Test (SPT), a miniature non-standard experimental device developed in the early 80s \cite{M81}. Its main attribute resides in the very small specimens employed (generally 8 mm diameter and 0.5 mm thickness), such that it may be considered a non-destructive experiment. The SPT has consistently proven to be a reliable tool for estimating the mechanical \cite{R09,G14} and creep \cite{DM09,J15} properties of metallic materials, as well as its environmentally-assisted degradation \cite{G15}. However, its capabilities in fracture toughness characterization are still a controversial subject. \\

A large experimental literature has appeared seeking to estimate the fracture toughness in metals by means of the SPT. Several authors \cite{MT87,B07,R13} have tried to accomplish this task by establishing a correlation with the so-called maximum biaxial strain, measured in the failure region of the SPT sample. Although good empirical correspondences have been found, results reveal a strong dependence on the material employed. Other schemes involve the use of neural networks to identify ductile damage parameters \cite{AK06} or energy-based approaches \cite{R09}. Recent research efforts have been mainly focused on the development of notched samples with the aim of increasing the attained constraint level \cite{J03,G15b}. The notch acts as a stress concentrator aiming to provide a triaxiality state closer to the standard fracture tests.\\

\begin{figure}[H]
\centering
\includegraphics[scale=0.6]{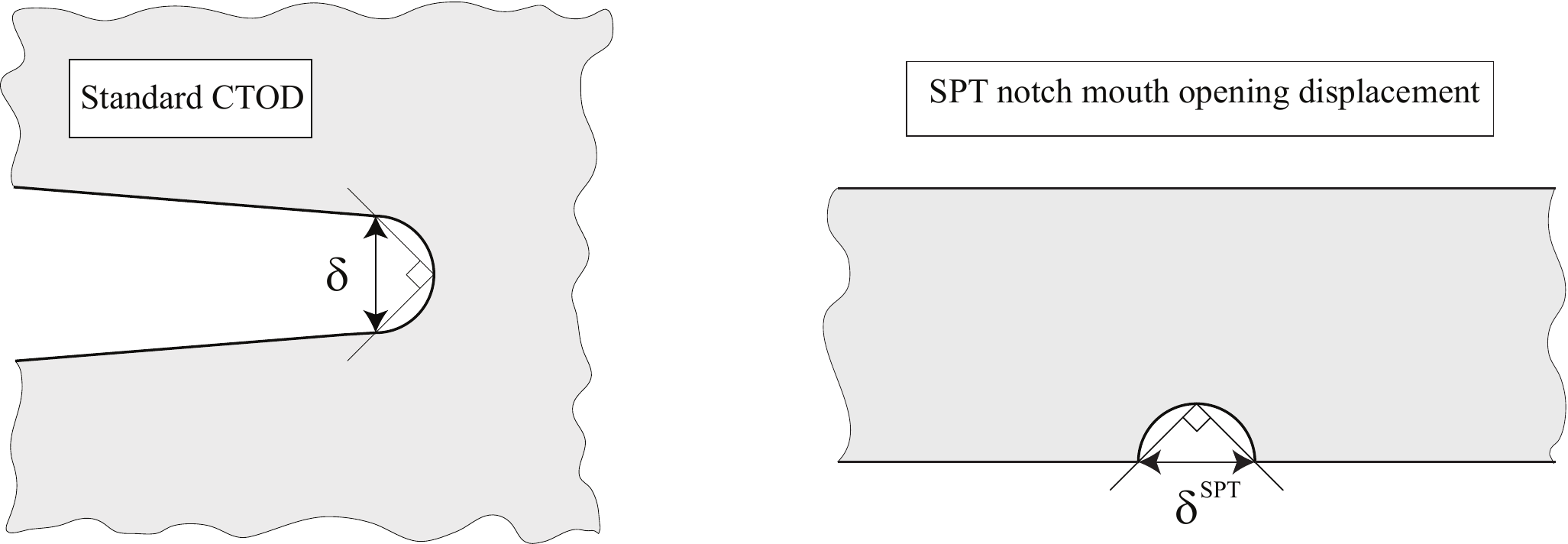}
\caption{Correlation between the standard CTOD in standard fracture tests and the notch mouth opening in the SPT}
\label{fig:CTODcomparison}
\end{figure}

In this work a hybrid numerical-experimental methodology for estimating the fracture toughness by means of the SPT is presented. The key objectives are: (i) to establish a procedure to classify industrial components as a function of their fracture resistance, and (ii) to set an appropriate correlation with standard tests, enabling a quantitative toughness assessment. A micromechanical damage model is employed to overcome the existing experimental shortcomings and enable the former objective, while the latter goal is facilitated by the use of the notch mouth opening displacement $\delta^{\mathrm{SPT}}$ as a fracture parameter, inspired by the standard crack tip opening displacement (CTOD) $\delta$ \cite{G15b}. As depicted in Fig. \ref{fig:CTODcomparison}, a parallelism can be established between the standard definition of the CTOD and the displacement of the notch faces in the SPT. The proposed methodology is comprehensively examined, with interrupted tests being performed to gain insight into the mechanisms behind cracking nucleation under SPT load conditions. Results obtained for two different grades of CrMoV steel are compared with standard fracture measurements and the outcome is thoroughly discussed.

\section{Materials and conventional characterization}
\label{Standard Testing}

As service conditions of hydrogen conversion reactors in the petrochemical industry are shifting to higher work temperatures and pressures, high thickness steel plates are being used in pressure vessels manufacturing and conventional 2.25Cr1Mo and 3Cr1Mo steels are progressively replaced by vanadium-modified low alloy steels such as 2.25Cr1MoV, 3Cr1MoV and 9Cr1MoV. In the present work, the structural integrity of a CrMoV steel welding joint is assessed through small scale test techniques by examining both base and weld metals. Thus, a 108 mm thick plate of 2.25Cr1Mo0.25V steel (SA 542 Grade D-Class 4) is employed for the base metal, which is subsequently normalized at 950$^{\circ}$C, quenched in water from 925$^{\circ}$C and tempered during 3 h. at 720$^{\circ}$C. The weld metal is obtained from a weld coupon of 1300 mm length and 600 mm width that is produced using a maximum gap of 30 mm by means of a submerged arc welding procedure using alternating current, a 4 mm diameter Thyssen Union S1 CrMo2V consumable and a heat input of 2.2 kJ/mm (29-32 V, 425-550 A and 45-55 cm/min); with an essential de-hydrogenation being performed immediately after welding. The chemical composition of the base (CrMoV1) and weld (CrMoV2) metals examined are shown in Table \ref{tab:Composition}.

\begin{table}[h]
\caption{Chemical composition of the CrMoV base (CrMoV1) and weld (CrMoV2) metals}
\centering
\begin{tabular}{c c c c c c c c} 
\hline
 & \% C & \% Si & \% Mn & \% Cr & \% Mo & \% V & \% Ni\\
 \hline
 CrMoV1 & 0.15 & 0.09 & 0.52 & 2.17 & 1.06 & 0.31 & 0.19\\
 CrMoV2 & 0.08 & - & - & 2.28 & 0.93 & 0.24 & 0.03\\
 \hline
\end{tabular}
\label{tab:Composition}
\end{table}

\subsection{Smooth tensile tests}

Three tensile tests per steel grade are performed following the ISO 6892-1:2009 standard. Smooth cylindrical bars are employed to mechanically characterize the behavior of both base and weld metals. The plastic behavior in the resulting stress-strain curves is fitted by means of Hollomon type power law:

\begin{equation}
\sigma=k \varepsilon^n_p
\end{equation}

\begin{figure}[H]
\centering
\includegraphics[scale=0.9]{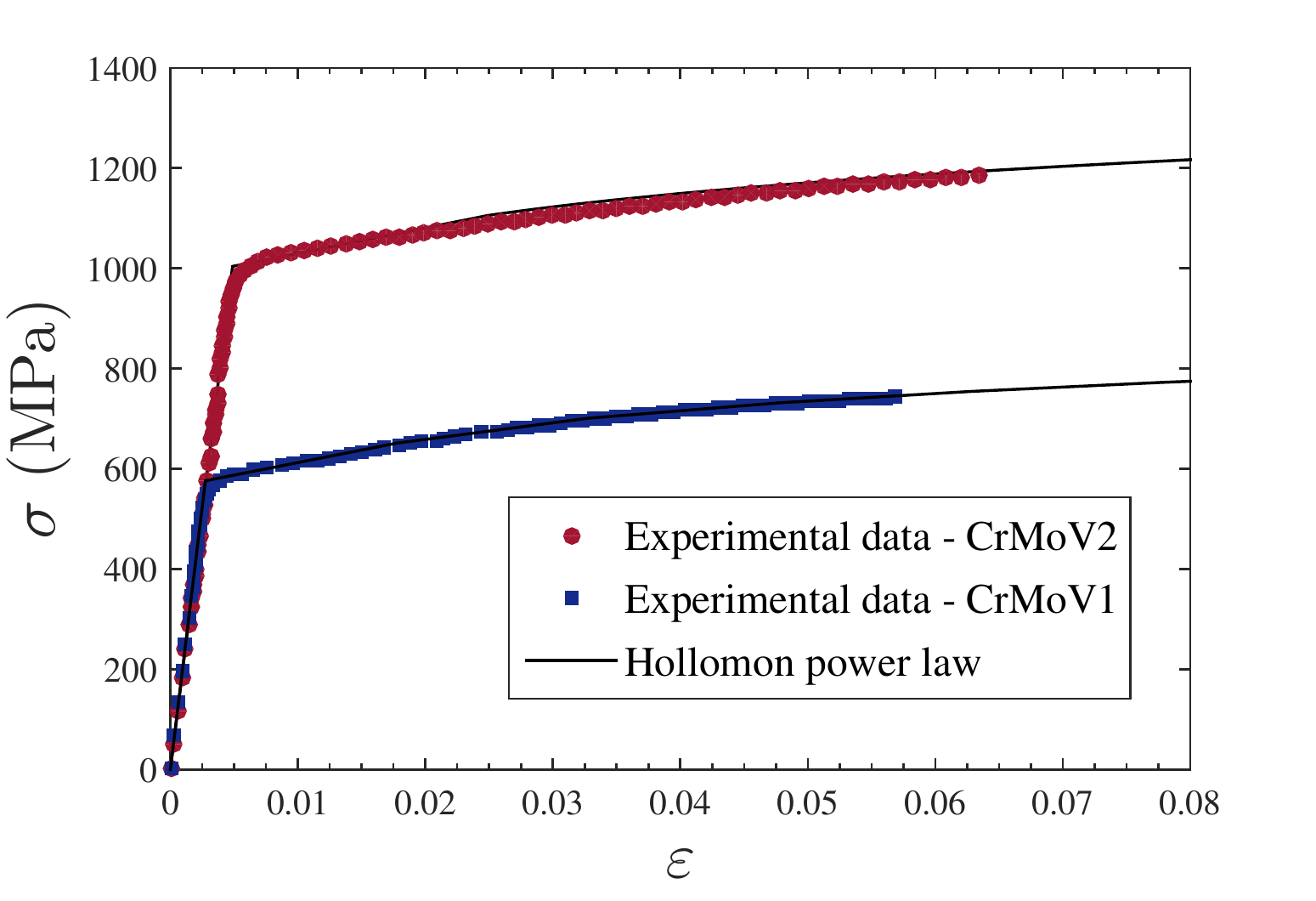}
\caption{Uniaxial stress strain curve for both base (CrMoV1) and weld (CrMoV2) metals}
\label{fig:SScurve}
\end{figure}

\noindent Where $\sigma$ is the uniaxial stress, $k$ is the strength coefficient, $\varepsilon_p$ is the equivalent plastic strain and $n$ is the strain hardening exponent. The experimental data for both CrMoV1 and CrMoV2, along with the power law fitting, are shown in Fig. \ref{fig:SScurve}.

The mechanical properties of both materials, particularly relevant for the finite strains finite element (FE) model, are summarized in Table \ref{tab:MechProperties}; where $E$ is Young's modulus, $\nu$ is the Poisson's ratio, $\sigma_Y$ is the yield stress and $\sigma_{UTS}$ is the ultimate tensile strength.

\begin{table}[h]
\caption{Mechanical properties}
\centering
\begin{tabular}{c c c c c c c} 
\hline
 & E (GPa) & $\,\,\,\, \nu \,\,\,\,$ & $\sigma_Y$ (MPa) & $\sigma_{UTS}$ (MPa) & $k$ (MPa) & $n$ \\
 \hline
 CrMoV1 & 200 & 0.3 & 595 & 711 & 1019 & 0.107 \\
 CrMoV2 & 236 & 0.3 & 1034 & 1121 & 1474 & 0.075 \\
 \hline
\end{tabular}
\label{tab:MechProperties}
\end{table}

\subsection{Notched tensile tests}

Uniaxial tensile tests are performed on circumferentially notched cylindrical bars to extract the micromechanical parameters employed in the ductile damage characterization of the base metal (see Section \ref{Finite element results}). Experiments are conducted following the ISO 6892-1:2009 standard, with the net diameter and the notch radius of the bar being equal to 5.26 mm and 1.16 mm, respectively. The vertical displacement is accurately measured as a function of the load through digital image correlation (DIC), a full-field optical technique used to capture displacement fields by comparing digital images of a specimen surface before and after deformation.

\subsection{Fracture tests}

\begin{figure}[H]
        \centering
        \begin{subfigure}[h]{0.49\textwidth}
                \centering
                \includegraphics[scale=0.38]{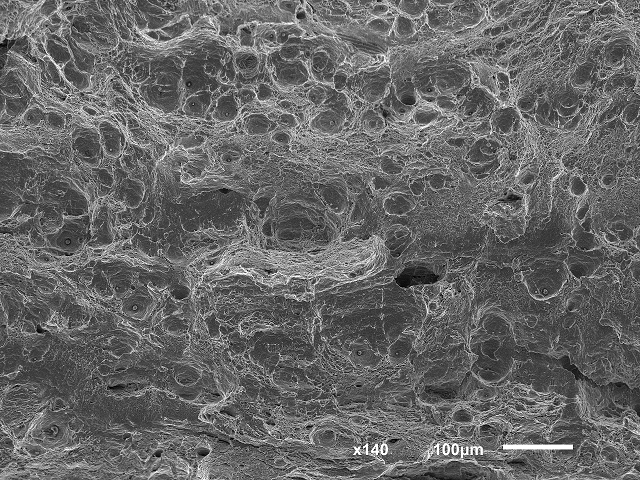}
                \caption{}
                \label{fig:FracCrMoV1}
        \end{subfigure}
        \begin{subfigure}[h]{0.49\textwidth}
                \centering
                \includegraphics[scale=0.38]{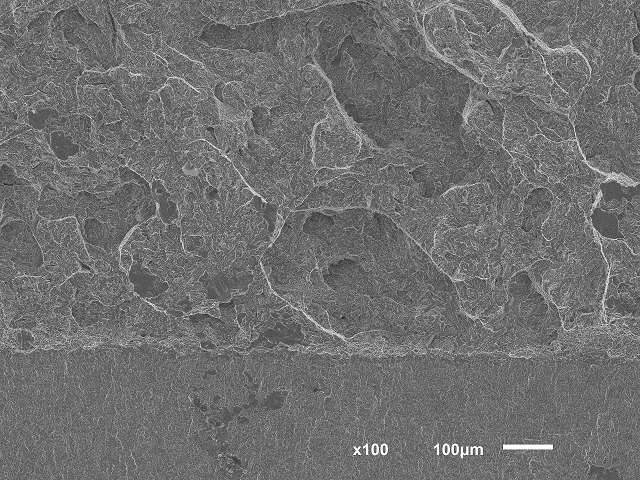}
                \caption{}
                \label{fig:FracCrMoV2}
        \end{subfigure}
       
        \caption{SEM fracture surface morphology: (a) Base metal (CrMoV 1) and (b) Weld metal (CrMoV 2)}\label{fig:FractureSurface}
\end{figure}

Fracture toughness tests are performed using single edge notched bend specimens, SE(B), with a crack length to width ratio: $a/W \approx 0.5$ (particularly, CrMoV1: $a_0=25.38$ mm and $W=49.99$ mm, while in CrMoV2: $a_0=22.1$ mm and $W=44.02$ mm), following the ASTM E1820 standard. Specimens are fatigue pre-cracked to the required nominal $a/W$ using a load ratio of 0.1. Results reveal significant differences between the base and the weld metals. Thus, the base metal (CrMoV 1) exhibits fully ductile behavior while brittle micromechanisms dominate fracture in the weld metal (CrMoV 2). Figure \ref{fig:FractureSurface} shows scanning electron microscope (SEM) fractographs revealing the main features observed in the fracture surface in both cases. Hence, Fig. \ref{fig:FracCrMoV1} shows a dimpled fracture surface, typical of microvoid coalescence, while in Fig. \ref{fig:FracCrMoV2} mainly brittle fracture is observed. 

Consequently, the single-specimen method (based on the use of the elastic unloading compliance technique) is used to determine the $\delta - \Delta a$ resistance curve of the CrMoV 1. The results obtained were thus corrected using the physical measure of the crack, determined at the end of each test through a suitable low magnification microscope. The value of $\delta$ for each unload is obtained after splitting up its elastic and plastic components:

\begin{equation}
\delta_i=\delta^{\mathrm{el}}_i+\delta^{\mathrm{pl}}_i
\end{equation}

\noindent Where the elastic component is obtained from the stress intensity factor, $K$:

\begin{equation}\label{Eq:deltaEl}
\delta^{\mathrm{el}}_i=\frac{K_i^2 (1-\nu^2)}{2 \sigma_Y E}
\end{equation}

With $K$ being estimated from the following expression, in accordance with the ASTM E1820 standard:

\begin{equation}\label{Eq:K1}
K_{(i)}=\left[ \frac{P_i S}{\left( B B_N \right)^{1/2} W^{3/2}} \right] f \left(a_i / W \right)
\end{equation}

\noindent where $P_i$ is the applied load, $S$ is the specimen span, $B$ and $B_N$ are the thickness and net thickness, respectively, and $f\left(a_i / W \right)$ is the configuration-dependent dimensionless function.

On the other hand, the plastic component of $\delta$ is computed through:

\begin{equation}
\delta^{\mathrm{pl}}_i=\frac{r_p (W-a_i) v_{\mathrm{pl}}}{r_p (W-a_i)+a_i}
\end{equation}

Where $v_{\mathrm{pl}}$ is the plastic component of the crack mouth opening displacement and the value of $r_p$ is given by the ASTM E1820 standard ($r_p$ = 0.44). Besides, the following power law:

\begin{equation}
\delta=C_1 \left( \Delta a \right)^{C_2}
\end{equation}

\noindent is employed to fit the experimental points $\delta_i-\Delta a_i$. On the other hand, in the case of the CrMoV 2 (brittle behavior), $\delta_{IC}$ is assessed by means of Eq. (\ref{Eq:deltaEl}).

\section{Small Punch Tests}
\label{SPT}

\begin{figure}[H]
\centering
\sbox{\measurebox}{%
  \begin{subfigure}[b]{.55\textwidth}
    {\includegraphics[width=\textwidth,height=5cm]{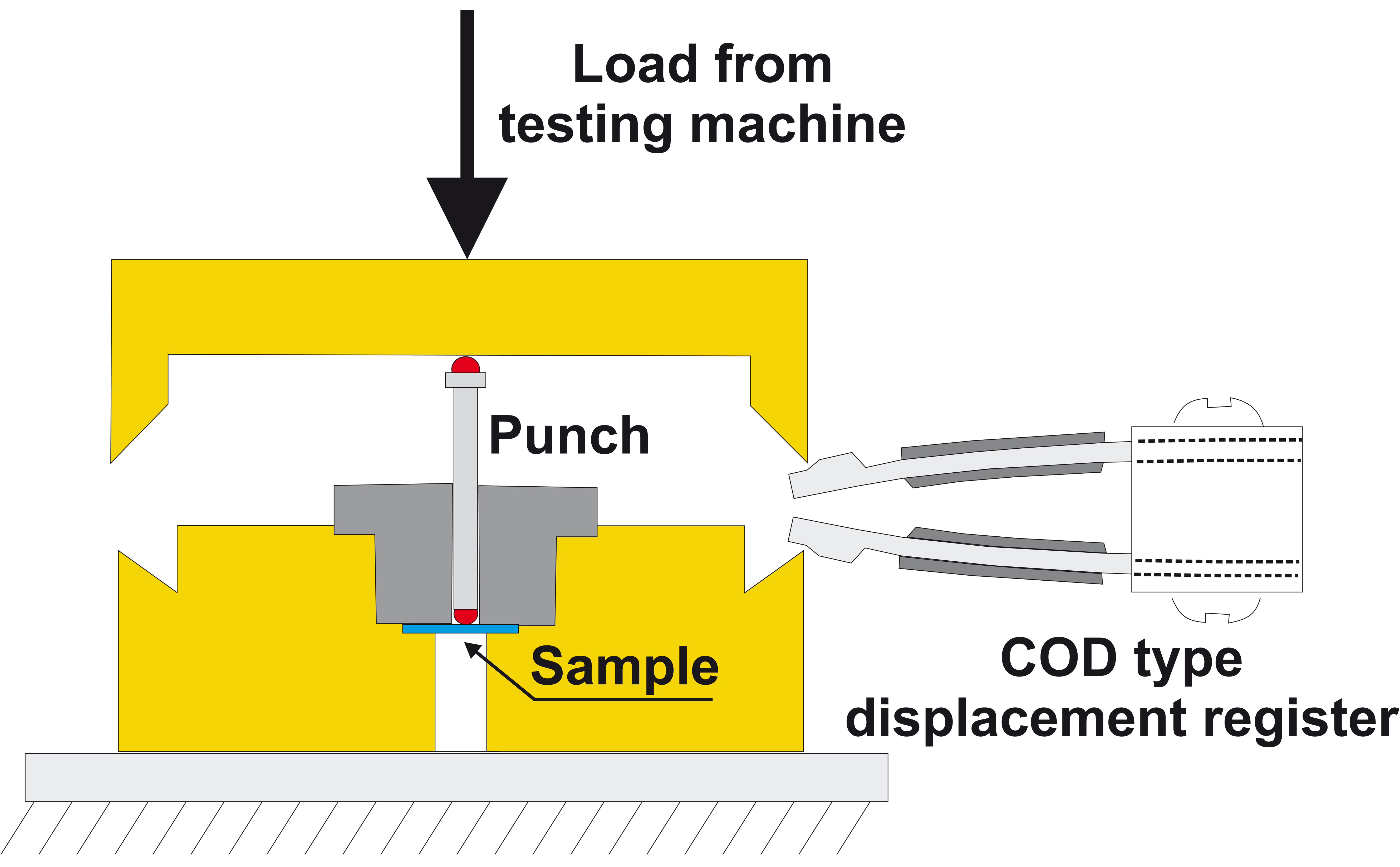}}
    \caption{}
    \label{fig:NotchedSPTA}
  \end{subfigure}}
\usebox{\measurebox}\qquad
\begin{subfigure}[b][\ht\measurebox][s]{.31\textwidth}
\centering
  {\includegraphics[width=\textwidth,height=2cm]{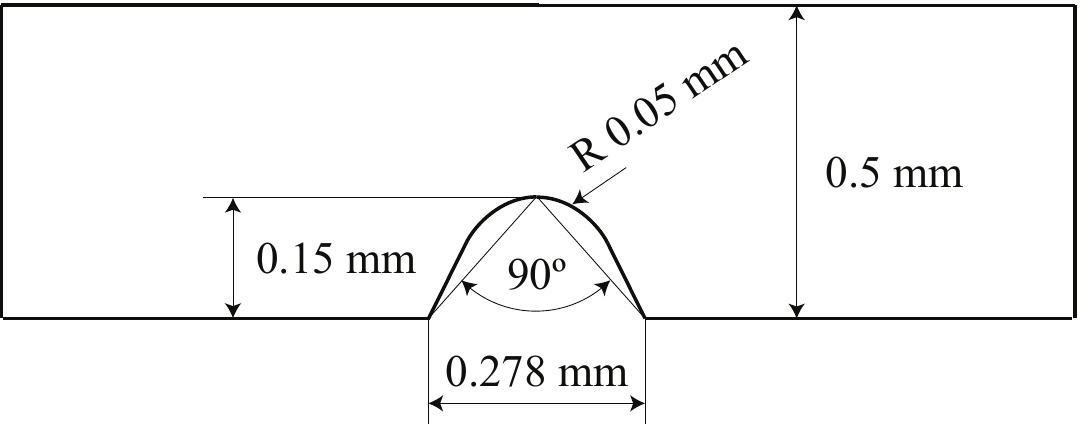}}
  \caption{}
  \label{fig:NotchedSPTB}
\vfill

  {\includegraphics[width=\textwidth,height=2cm]{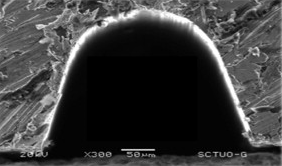}}
  \caption{}
  \label{fig:NotchedSPTC}
\end{subfigure}
\caption{Notched SPT specimen. (a) Experimental device, (b) notch geometry and (c) SEM image of the notch}
\label{fig:NotchedSPT}
\end{figure}

SPT samples of 10 mm length and 10 mm width with a thickness of $t=0.5 \pm 0.01$ mm are obtained by means of a precision metallographic cutting machine. A blind longitudinal notch is inserted through micro-machining with the aim of ensuring a uniform shape and depth along the entire specimen length. The notch dimensions and its location within the specimen are depicted in Fig. \ref{fig:NotchedSPTB}. Following the standard definition of the CTOD (displacement at the intersection of a $90^{\circ}$ vertex with the crack flanks), its SPT counterpart ($\delta^{\mathrm{SPT}}$) is obtained by measuring, through SEM and image analysis software, the notch mouth opening displacement \cite{G15b}.  

The small punch tests are performed using a special device (outlined in Fig. \ref{fig:NotchedSPTA})  that is attached to a universal testing machine. The entire contour of the sample is firmly pressed between two dies while the load is applied to the center of the specimen by means of a 2.5 mm hemispherical diameter punch. A free-standing extensometer is coupled to the experimental device to accurately measure the punch displacement. A displacement rate of 0.2 mm/min is used and lubrication is employed to minimize the effects of friction \cite{G14,G15b}.

\section{Numerical modelization}
\label{Finite element results}

\begin{figure}[H]
\centering
\includegraphics[scale=0.9]{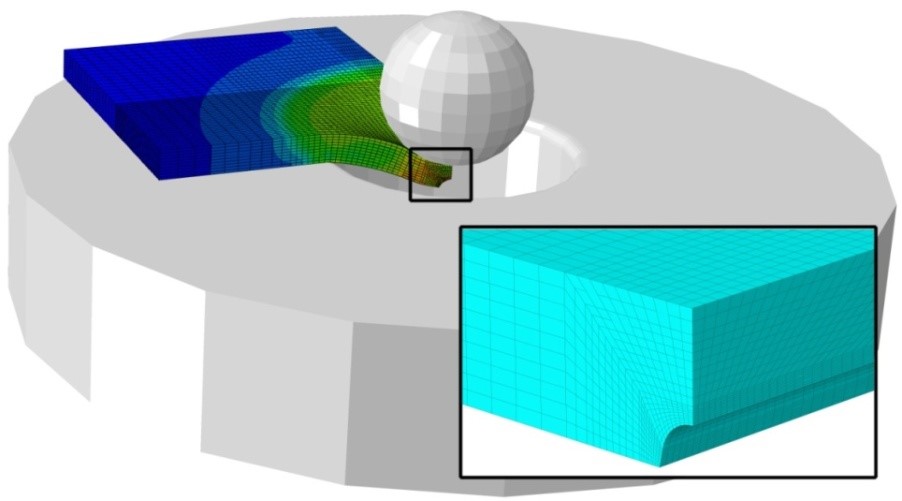}
\caption{Numerical model of the notched SPT specimen (the fixer is left out for visualization purposes).}
\label{fig:FEmesh}
\end{figure}

A finite element model of the SPT is developed using the commercial software ABAQUS/Standard version 6.13. Attending to the specimen geometry and test setup, a 3-D approach is adopted. Due to symmetry, only one quarter of the specimen is modeled by means of 44400 8-node linear brick elements (C3D8). A more refined mesh is used near the notch, where a minimum element length of 0.025 mm is employed after the corresponding sensitivity study. The lower matrix, the fixer and the punch are modeled as rigid bodies and their degrees of freedom are restricted except for the vertical displacement of the punch. The friction coefficient was set to $\mu=0.1$, which is a common value for steel-to-steel contact in the presence of lubrication. The scheme of the model and the mesh employed are shown in Figure \ref{fig:FEmesh}.

Results obtained from tensile tests are used to characterize the elasto-plastic behavior. The influence of nucleation, growth and coalescence of microvoids is modeled by means of the well-known Gurson-Tvergaard-Needleman (GTN) model \cite{G75,TN84}. In this model, the yield function is defined by:

\begin{equation}
\Phi \left( q, \sigma_m , \sigma , f \right)= \left( \frac{q}{\sigma} \right)^2 + 2 q_1 f^* \textnormal{cosh} \left( \frac{3 q_2 \sigma_m}{2 \sigma} \right) - \left(1+ q_3 {f^*}^2\right) =0
\end{equation}

\noindent Where $f$ is the void volume fraction, $\sigma_m$ is the mean normal stress, $q$ is the conventional Von Mises equivalent stress, $\sigma$ is the flow stress of the matrix material and $q_1$, $q_2$ and $q_3$ are fitting parameters introduced by Tvergaard \cite{T82}. The modified void volume fraction $f^*$ was introduced by Tvergaard and Needleman \cite{TN84} to predict the rapid loss in strength that accompanies void coalescence, and it is given by:

\[ f^* =
  \begin{cases}
    f & \quad \text{for } f \le f_c \\
    f_c+\frac{f_u^*-f_c}{f_f-f_c} \left(f-f_c \right) & \quad \text{for } f > f_c\\
  \end{cases}
\]

Where $f_c$ is the critical void volume fraction, $f_f$ is the void volume fraction at final fracture and $f^*_u=1/q_1$ is the ultimate void volume fraction. Thus, the evolution law for the void volume fraction is given in the model by an expression of the form:

\begin{equation}
\dot{f}=\dot{f}_{\mathrm{growth}}+\dot{f}_{\mathrm{nucleation}}
\end{equation}

According to Chu and Needleman \cite{CN80} the nucleation rate is assumed to follow a Gaussian distribution, that is:

\begin{equation}
\dot{f}_{\mathrm{nucleation}}=A \dot{\bar{\varepsilon}}^p
\end{equation}

Where $\dot{\bar{\varepsilon}}^p$ is the equivalent plastic strain rate, and:

\begin{equation}
A= \frac{f_n}{S_n \sqrt{2 \pi}} \textnormal{exp} \left(-\frac{1}{2} \left(\frac{\bar{\varepsilon}^p-\varepsilon_n}{S_n} \right)^2 \right)
\end{equation}

Being $\varepsilon_n$ the mean strain, $S_n$ the standard deviation and $f_n$ the void volume fraction of nucleating particles. 

The GTN model is implemented in ABAQUS by means of a UMAT subroutine, where the consistent tangent moduli is computed through an implicit Euler backward algorithm, as proposed by Zhang \cite{Z95}. Following a common procedure in the literature, GTN parameters are obtained by assuming $q_1=1.5$, $q_2=1.0$, $q_3=2.25$ \cite{T81}, $\varepsilon_n=0.3$, $S_n=0.1$ \cite{CN80} and calibrating $f_0$, $f_n$, $f_c$ and $f_f$ with experiments through a top-down approach \cite{BL10}. The initial void volume fraction $f_0$ is assumed to be equal to 0 for the CrMoV steels analyzed in this work as it is associated with the volume fraction of intermetallic particles, and $f_n$, $f_c$ and $f_f$ are obtained by matching numerical simulations and the load-displacement curve (LDC) from uniaxial testing of notched round bars (see Section \ref{Standard Testing}). The displacement corresponds to the relative vertical displacement between two points equidistantly located 1.76 mm from the center of the bar. Because of double symmetry only one quarter of the notched tensile specimen is modeled through 1516 8-node quadrilateral axisymmetric elements (CAX8).\\

\begin{figure}[H]
\makebox[\linewidth][c]{%
        \begin{subfigure}[b]{0.74\textwidth}
                \centering
                \includegraphics[scale=0.45]{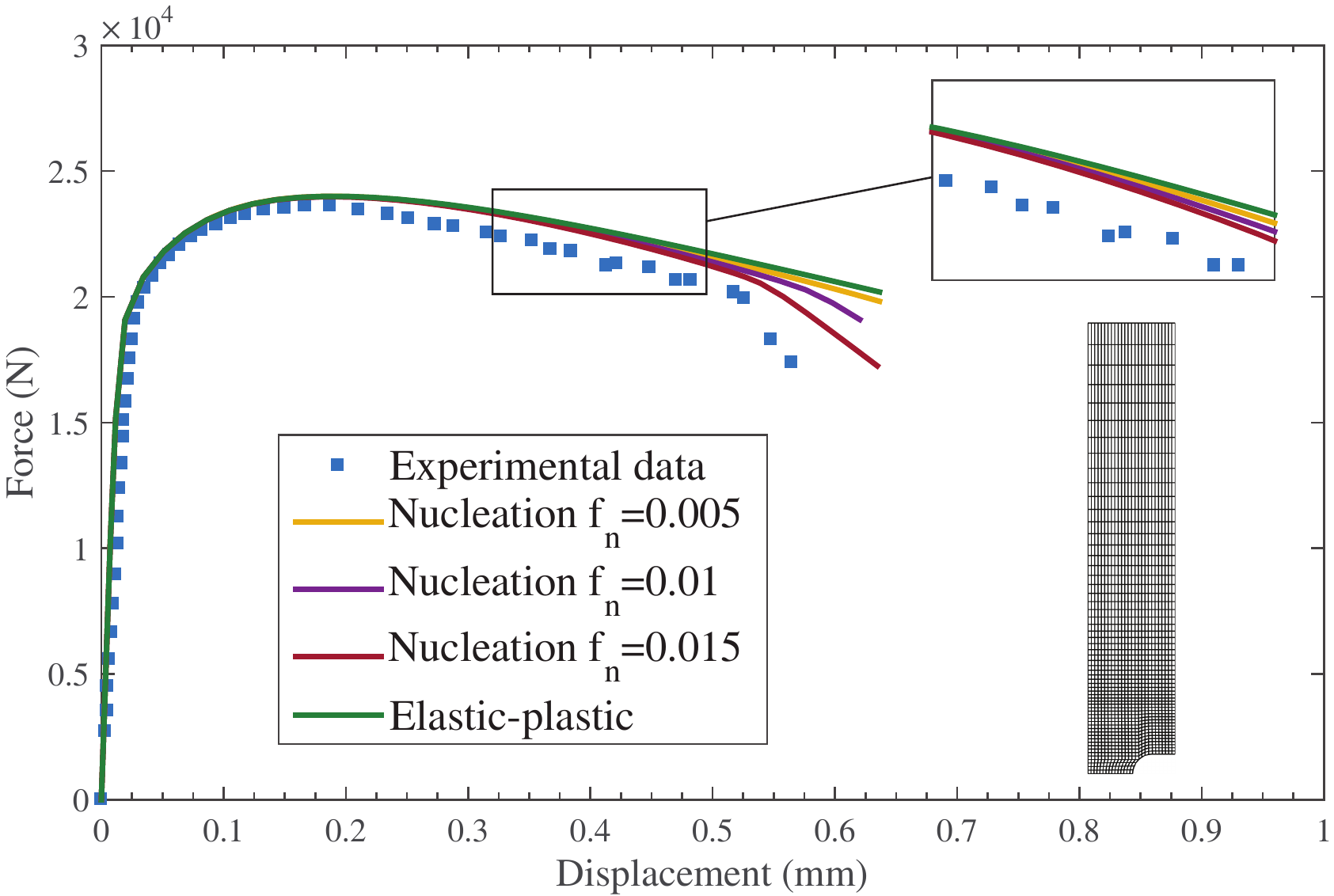}
                \caption{}
                \label{fig:TopDownA}
        \end{subfigure}
        \begin{subfigure}[b]{0.6\textwidth}
                \raggedleft
                \includegraphics[scale=0.45]{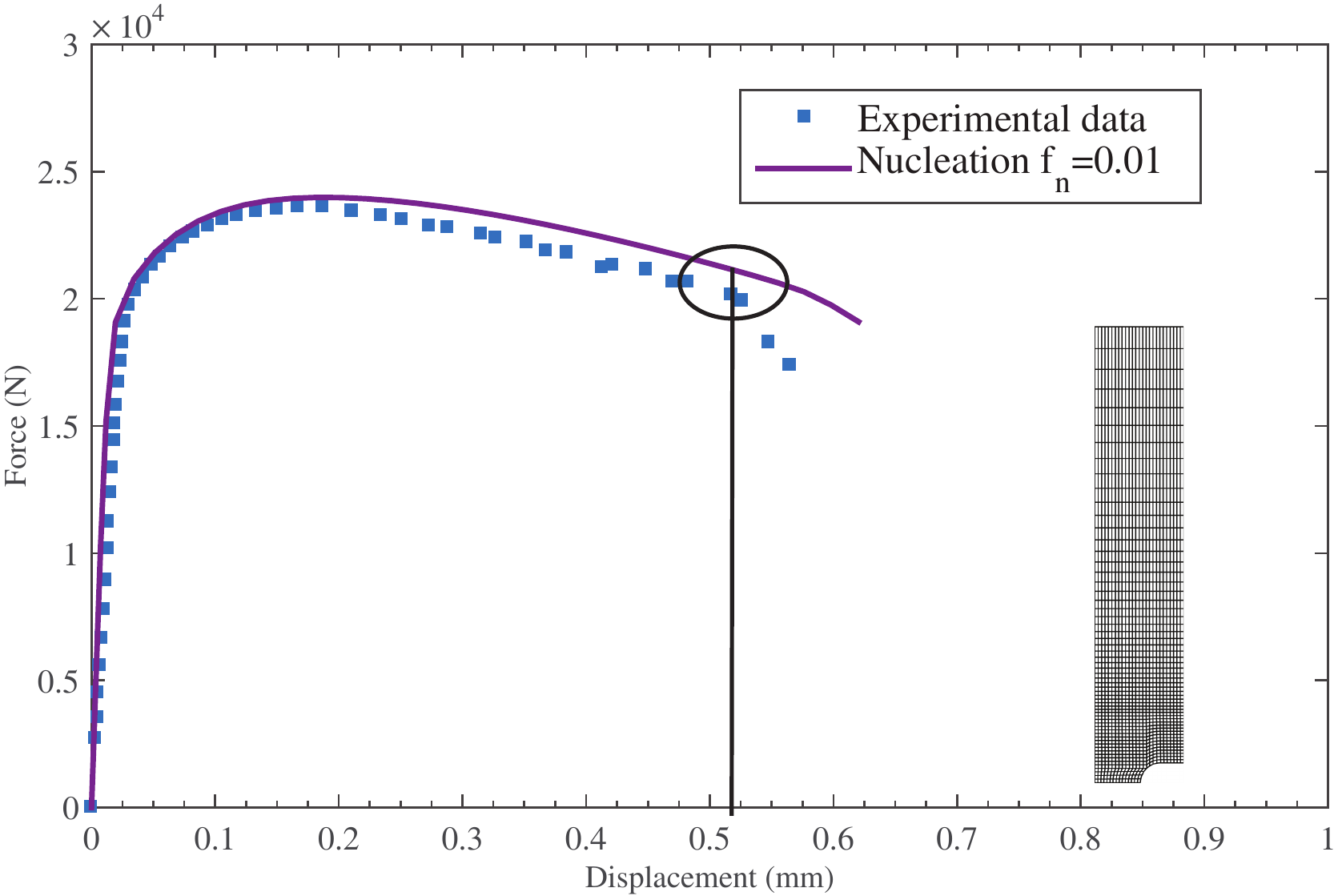}
                \caption{}
                \label{fig:TopDownB}
        \end{subfigure}}

\makebox[\linewidth][c]{%
        \begin{subfigure}[b]{0.74\textwidth}
                \centering
                \includegraphics[scale=0.45]{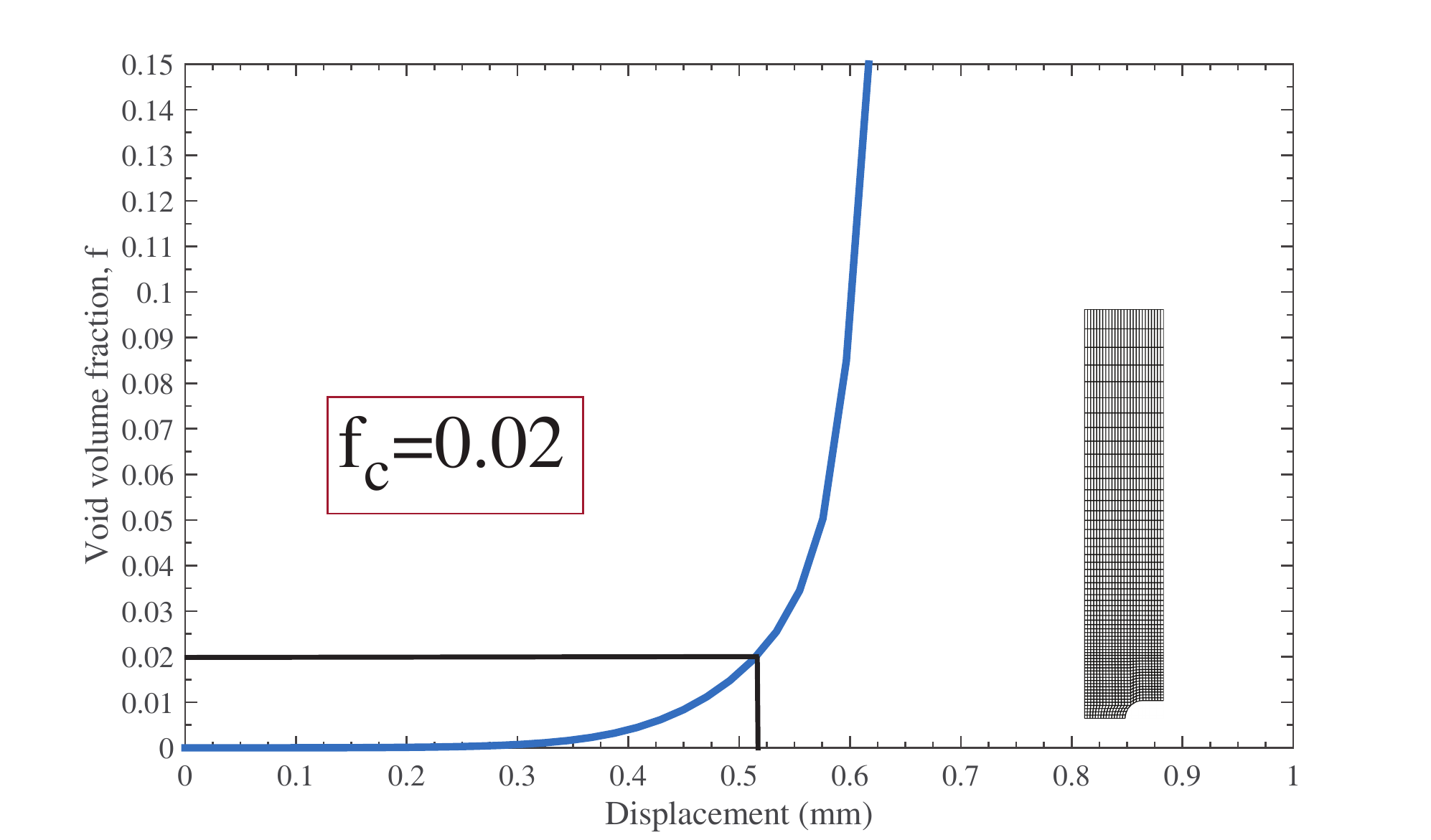}
                \caption{}
                \label{fig:TopDownC}
        \end{subfigure}
        \begin{subfigure}[b]{0.5\textwidth}
                \raggedleft
                \includegraphics[scale=0.45]{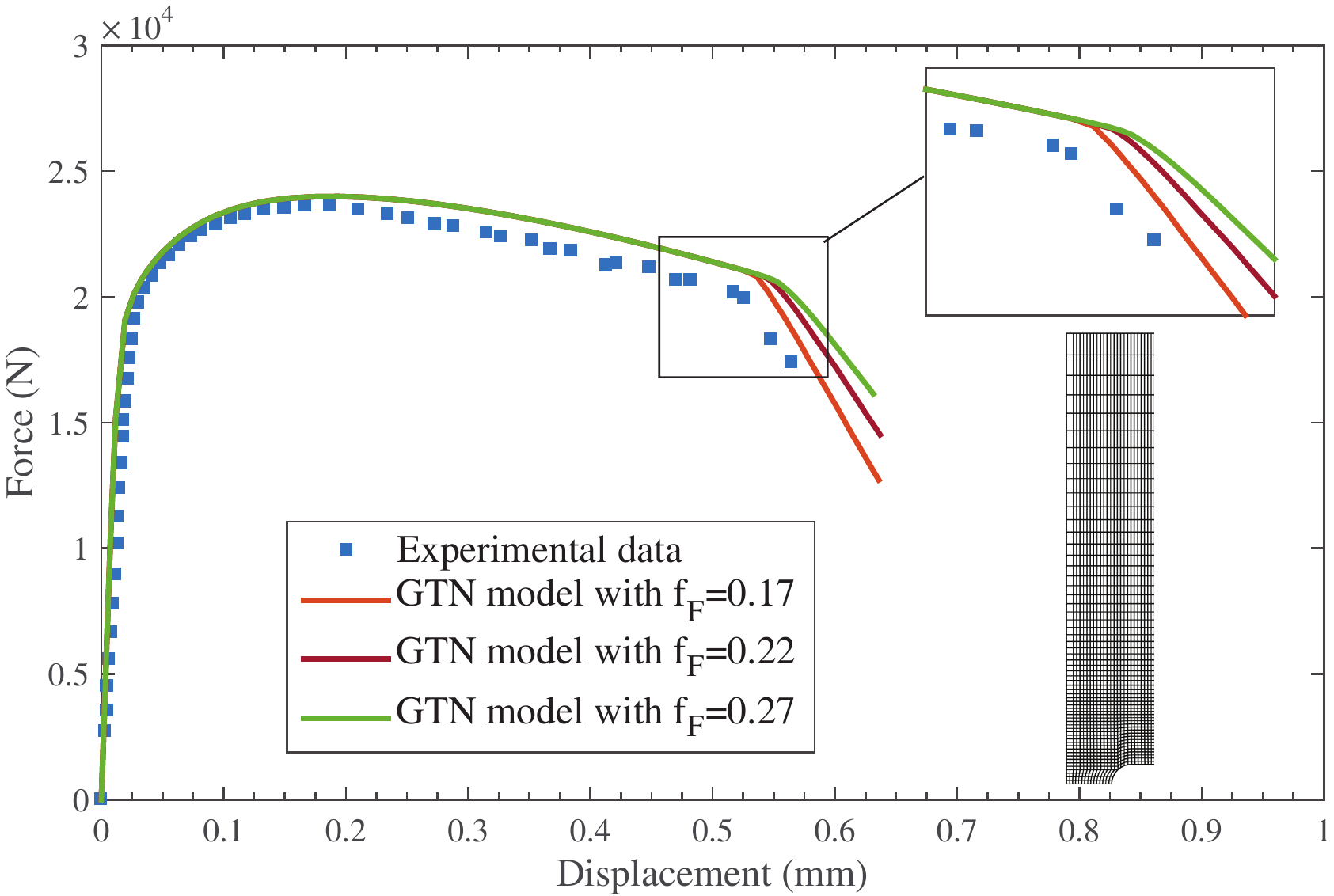}
                \caption{}
                \label{fig:TopDownD}
        \end{subfigure}
        }       
        \caption{Outline of the top-down approach: (a) Experimental data and numerical predictions for different $f_n$, (b) identification of the sudden load drop associated with void coalescence, (c) void volume fraction in the center of the specimen versus displacement for the chosen value of  $f_n$, (d) numerical damage simulation for different $f_f$.}\label{fig:TopDown}
\end{figure}

Fig. \ref{fig:TopDown} shows an outline of the followed methodology to determine the GTN parameters in the CrMoV1 case. As shown in Fig. \ref{fig:TopDownA}, the void volume fraction of nucleating particles $f_n$ is obtained by correlating the experimental data with the numerical results obtained omitting the failure criterion. Afterwards, Figs. \ref{fig:TopDownB} and \ref{fig:TopDownC}, the critical void volume fraction $f_c$ is identified by assuming that it corresponds with the initiation of void coalescence \cite{TN84}. Finally, the slope of the LDC once the load carrying capacity decreases drastically determines the value of $f_f$ (Fig. \ref{fig:TopDownD}).

Damage parameters obtained for the base metal (CrMoV1) by means of the aforementioned methodology are displayed in Table \ref{tab:GursonParameters}. It is not possible to compare the calibrated GTN parameters with previous works in the literature since, to the authors' knowledge, ductile damage characterization of CrMoV steel through the GTN model has not been analyzed before. However, similar values have been obtained in steels with a similar microstructure \cite{SZ97,S97}. 

\begin{table}[H]
\caption{Ductile damage modeling parameters (GTN model) of the base metal obtained from a notched tensile test through a top-down approach}
\centering
\begin{tabular}{c c c c c c c c c c} 
\hline
 & $q_1$ & $q_2$ & $q_3$ & $f_0$ & $\varepsilon_n$ & $S_n$ & $f_n$ & $f_c$ & $f_f$ \\
 \hline
 CrMoV1 & 1.5 & 1.0 & 2.25 & 0 & 0.3 & 0.1 & 0.01 & 0.02 & 0.22 \\
 \hline
\end{tabular}
\label{tab:GursonParameters}
\end{table}

GTN parameters can also be obtained from the load-displacement curve of the SPT (see, e.g., \cite{P09,C10}), but using notched tensile specimens enables us to clearly establish the location of the onset of damage and accurately measure the displacement through the DIC technique. Nonetheless, the aforementioned procedure cannot be employed in the weld metal case (CrMoV2) as brittle failure mechanisms are observed in standardized tests. A ductile damage model may, however, still be employed to characterize the weld metal response in the SPT, as the idiosyncrasy of the experiment favors failure through nucleation, growth and coalescence of microvoids. Hence, in a similar way as \cite{C07}, CrMoV2 GTN parameters (Table \ref{tab:GursonParameters1}) are obtained by fitting through trial and error the experimental SPT curve.

\begin{table}[H]
\caption{Ductile damage modeling parameters (GTN model) of the weld metal obtained by fitting the SPT load-displacement curve}
\centering
\begin{tabular}{c c c c c c c c c c} 
\hline
 & $q_1$ & $q_2$ & $q_3$ & $f_0$ & $\varepsilon_n$ & $S_n$ & $f_n$ & $f_c$ & $f_f$ \\
 \hline
 CrMoV2 & 1.5 & 1.0 & 2.25 & 0 & 0.3 & 0.1 & 0.035 & 0.045 & 0.24 \\
 \hline
\end{tabular}
\label{tab:GursonParameters1}
\end{table}

\begin{figure}[H]
\centering
\includegraphics[scale=0.8]{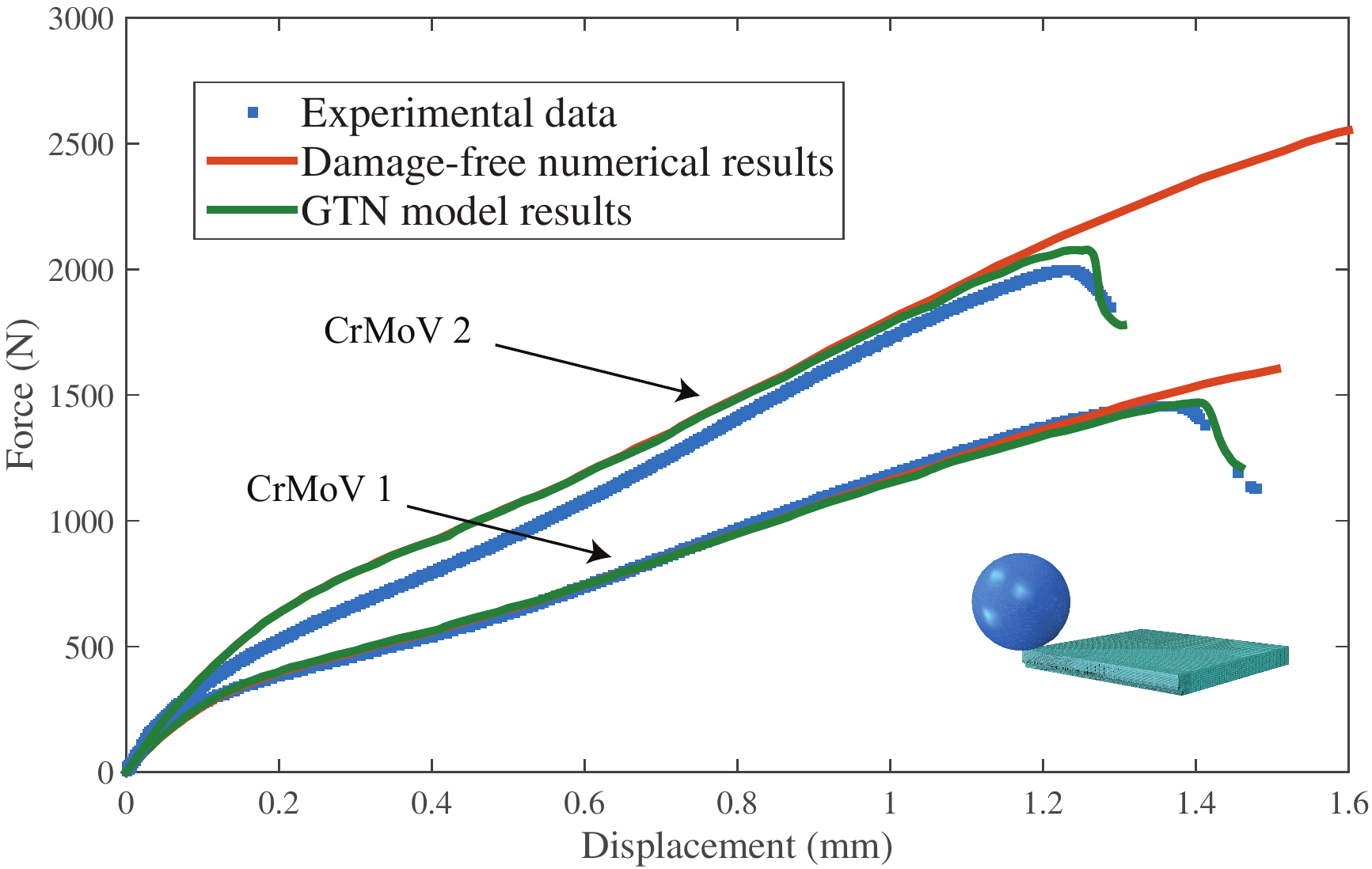}
\caption{SPT experimental and numerical (with and without damage) load-displacement curves}
\label{fig:NumResults}
\end{figure}

Once the model parameters have been obtained for both base and weld metals; nucleation, growth and coalescence of microvoids are incorporated in the SPT finite element framework. The results obtained are shown in Fig. \ref{fig:NumResults}, along with the experimental data and the conventional elasto-plastic predictions. The trends depicted in Fig. \ref{fig:NumResults} are consistent with the tensile properties of Table \ref{tab:MechProperties}, with the weld metal attaining a higher maximum load. Results reveal a good agreement between experimental and numerical damage-enhanced curves, proving the good performance of the top-down methodology employed to estimate the GTN parameters of the base metal. Nevertheless, the brittle behavior observed in conventional fracture mechanics testing of the weld metal reveals that damage modelization by means of parameters obtained from the SPT curve must be performed with caution. Thus, while a microvoid-based model may accurately capture the material response observed in the SPT experiments, higher stress triaxiality conditions may alter the hierarchy of mechanisms governing crack propagation.

As the nature of the experiment hinders the observation of the critical CTOD associated with the onset of crack growth, the numerical model plays a fundamental role enabling its accurate identification.

\section{Results}
\label{Results}

\begin{figure}[H]
        \centering
        \begin{subfigure}[h]{0.49\textwidth}
                \centering
                \includegraphics[scale=0.52]{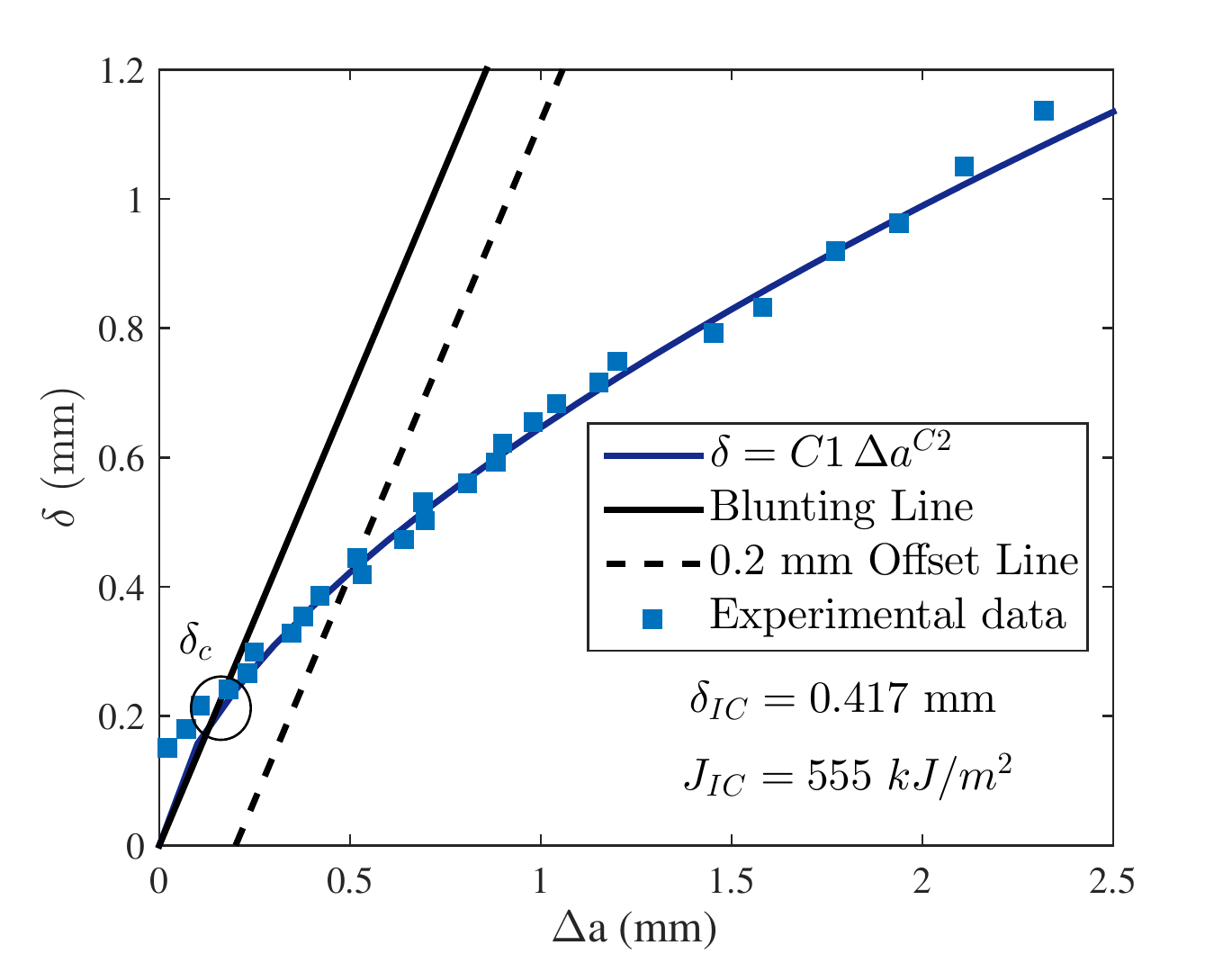}
                \caption{}
                \label{fig:FracStdCrMoV1}
        \end{subfigure}
        \begin{subfigure}[h]{0.49\textwidth}
                \centering
                \includegraphics[scale=0.52]{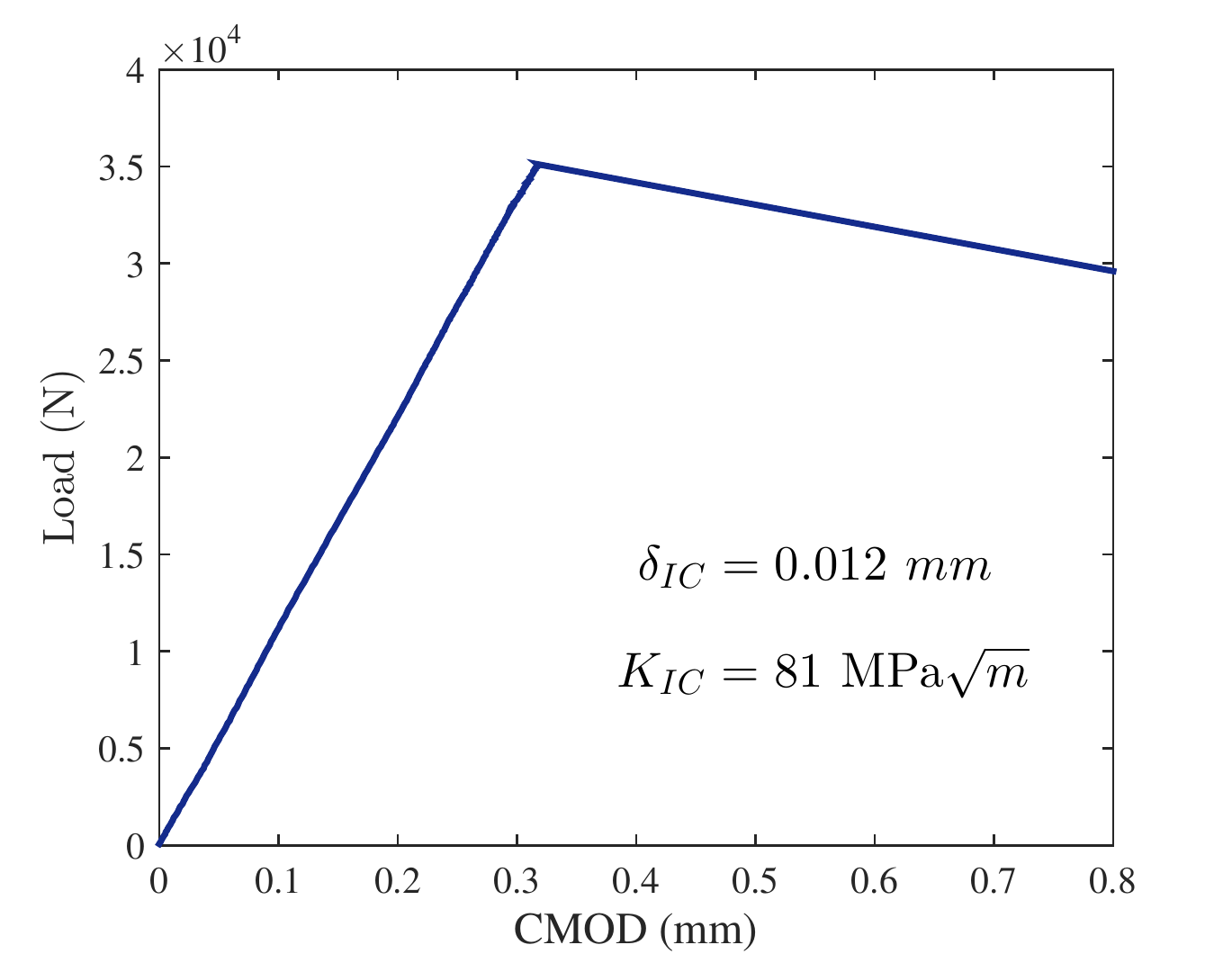}
                \caption{}
                \label{fig:FracStdCrMoV2}
        \end{subfigure}
       
        \caption{Conventional fracture characterization: (a) $\delta-\Delta a$ curve for CrMoV1 and (b) load-CMOD curve for CrMoV2}\label{fig:FractureStandard}
\end{figure}

Crack opening displacement measurements recorded in the standard fracture tests described in Section \ref{Standard Testing} are shown in Fig. \ref{fig:FractureStandard}. Thus, Fig. \ref{fig:FracStdCrMoV1} shows the CTOD variation as a function of crack growth in the base metal, while the load versus CMOD curve of the weld metal is plotted in Fig. \ref{fig:FracStdCrMoV2}. Critical toughness parameters ($\delta_{IC}$ and $J_{IC}$ or $K_{IC}$) are identified following the ASTM E1820 standard and shown in Fig. \ref{fig:FractureStandard}. Moreover, as crack growth cannot be estimated in the SPT, a critical measure of the CTOD $\delta_c$ is defined at the onset of crack propagation for comparison purposes. As shown in Fig. \ref{fig:FracStdCrMoV1}, $\delta_c$ is identified in the base metal as the crack tip opening displacement when the blunting line separates from the $\delta-\Delta a$ curve. On the other hand, $\delta_c$ equals the critical CTOD $\delta_{IC}$ in the weld metal, as a consequence of the unstable brittle fracture observed.

\begin{figure}[H]
        \centering
        \begin{subfigure}[h]{0.9\textwidth}
                \centering
                \includegraphics[scale=0.65]{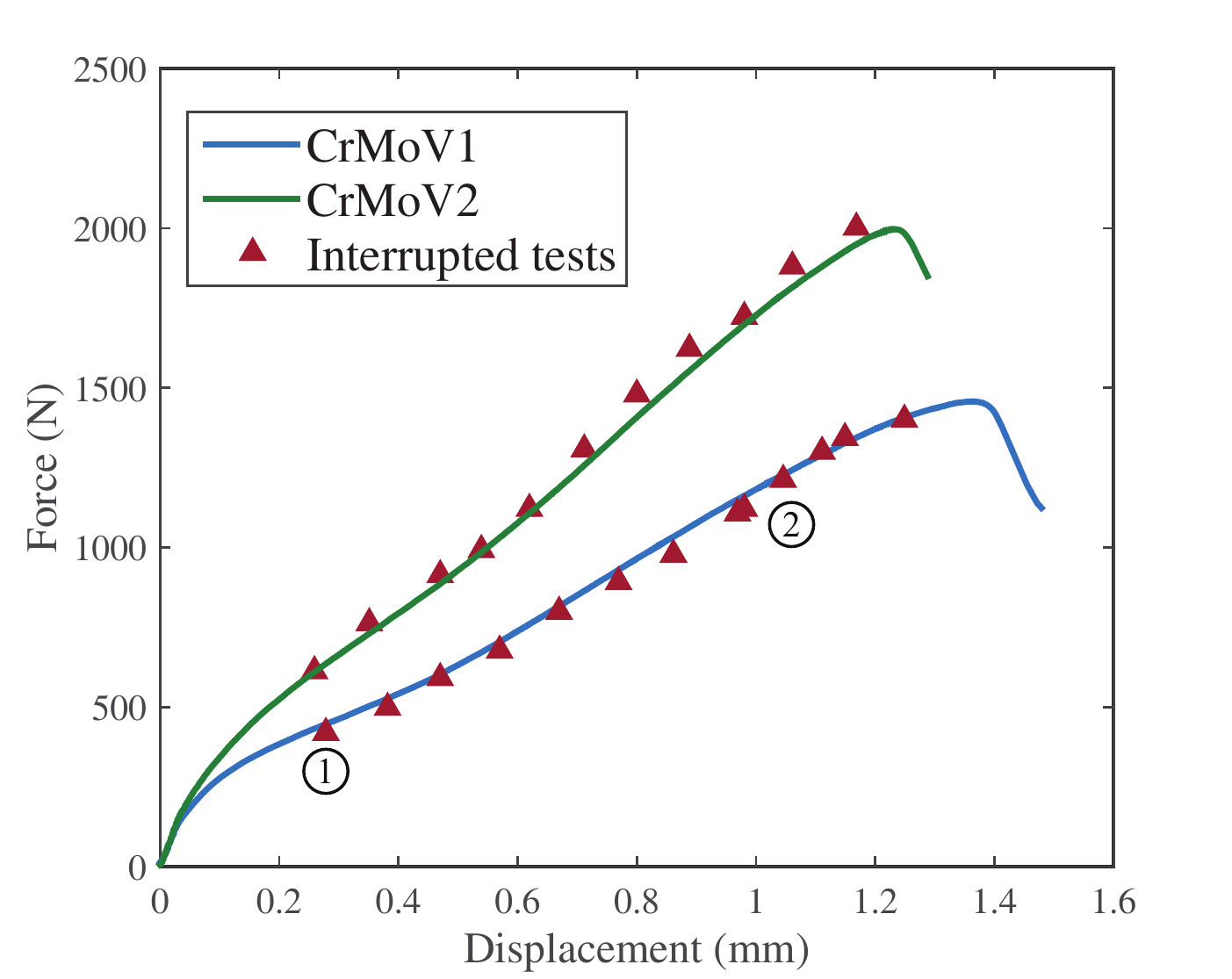}
                \caption{}
                \label{fig:LDCresults}
        \end{subfigure}\bigskip
        \par\bigskip
        \begin{subfigure}[h]{0.9\textwidth}
                \centering
                \includegraphics[scale=0.4]{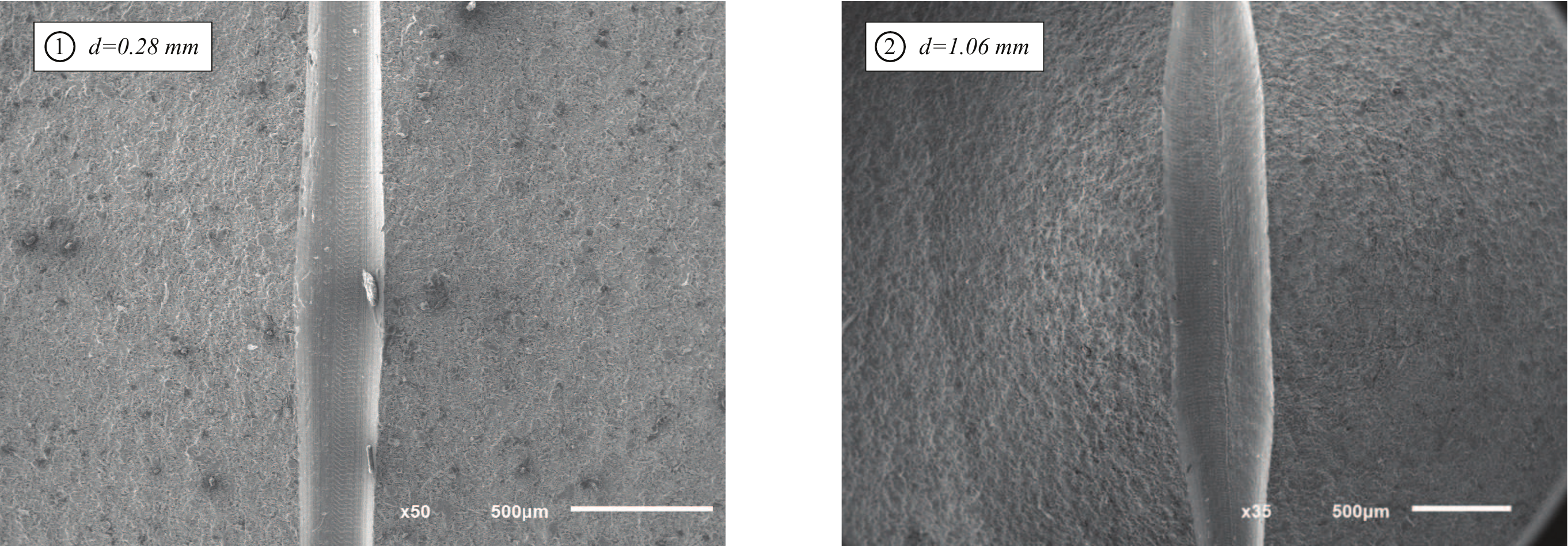}
                \caption{}
                \label{fig:SEMimages}
        \end{subfigure}
       
        \caption{Experimental SPT results: (a) Load-displacement curves, with the interrupted tests denoted by symbols and (b) SEM images of the notch at several punch displacement levels.}\label{fig:SPTresults}
\end{figure}

Experimental load-displacement curves obtained in the SPT are shown in Fig. \ref{fig:LDCresults}, where symbols denote the punch displacement levels at which the experiments have been interrupted. As shown in the figure, numerous tests are interrupted with the aim of physically measuring the notch mouth opening displacement $\delta^{\mathrm{SPT}}$ at several load levels. Micrographs corresponding to two particular punch displacement levels ($d=0.28$ and $d=1.06$ mm) are shown in Fig \ref{fig:SEMimages} as representative examples. SEM characterization of the interrupted specimens also enables to visualize cracks and gain insight into the damage mechanisms taking place.

\begin{figure}[H]
\centering
\includegraphics[scale=0.9]{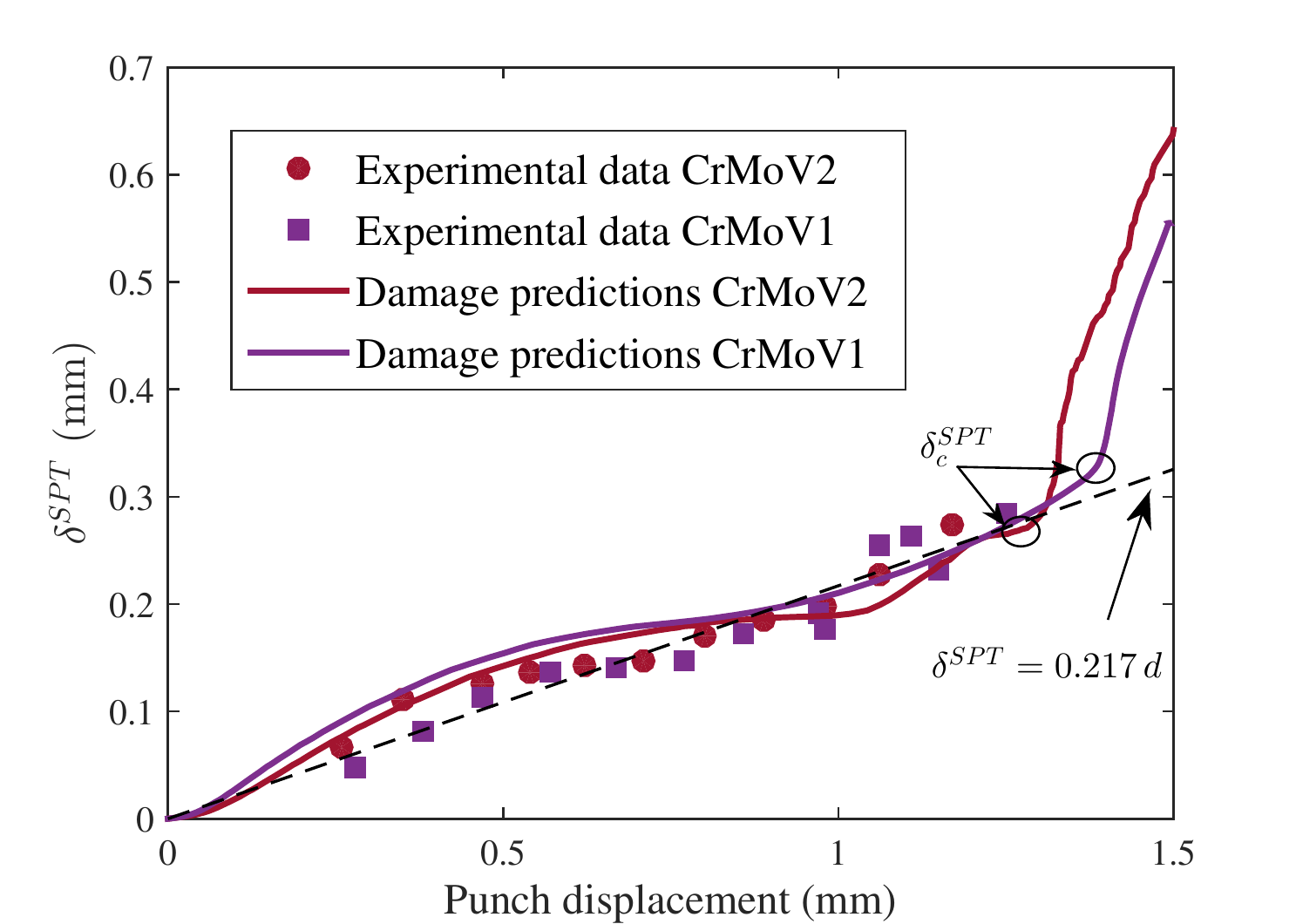}
\caption{Experimental and numerical notch mouth opening displacement versus punch displacement curves.}
\label{fig:CTODresults}
\end{figure}

Notch mouth opening measurements at each interrupted test are shown in Fig. \ref{fig:CTODresults} as a function of the punch displacement. Fig. \ref{fig:CTODresults} also shows the damage-enhanced numerical predictions for both the base metal (CrMoV1) and the weld metal (CrMoV2); where $\delta^{\mathrm{SPT}}$ is measured as the displacement of the notch faces, mimicking the experimental procedure.

A good agreement is observed between numerical and experimental predictions of notch mouth opening displacement in the SPT. Particularly, the results shown in Fig. \ref{fig:CTODresults} reveal two important features; on one hand, a linear relation can be easily observed between $\delta^{\mathrm{SPT}}$ measurements and the punch displacement $d$. This correlation that appears to show little sensitivity to material properties, is depicted in Fig. \ref{fig:CTODresults} by means of a dashed line and can be expressed as:

\begin{equation}
\delta^{\mathrm{SPT}}=0.217 \cdot d
\end{equation}

On the other hand, numerical damage predictions reveal a steep increase in $\delta^{\mathrm{SPT}}$ as the punch displacement approaches the maximum load. Hence, the finite element model reflects the rapid softening that takes place during void coalescence, allowing us to easily distinguish the onset of failure. This numerically-estimated critical point is identified as the notch mouth opening displacement at crack initiation $\delta^{\mathrm{SPT}}_c$. Thus, by means of the present combined numerical-experimental methodology it is possible to directly compare an SPT-based fracture toughness parameter with an equivalent measurement in conventional fracture experiments. Results obtained for the two steels considered in the present study, along with the standard test measurements, are shown in Table \ref{tab:CTOD}.

\begin{table}[H]
\caption{Critical tip displacement measurements at the initiation of crack growth}
\centering
\begin{tabular}{c c c} 
\hline
 & Standard Test $\delta_{c}$ (mm) & Small Punch Test $\delta^{\mathrm{SPT}}_c$ (mm)\\
 \hline
 CrMoV1 & 0.214 & 0.32 \\
 CrMoV2 & 0.012 & 0.26 \\
 \hline
\end{tabular}
\label{tab:CTOD}
\end{table}

In both steels the critical value measured in the standard tests is lower than its SPT counterpart. This could be expected as the constraint level is much higher in the normalized tests (approaching plane strain state), leading to a very conservative value of the fracture toughness.

\begin{figure}[H]
\centering
\includegraphics[scale=0.35]{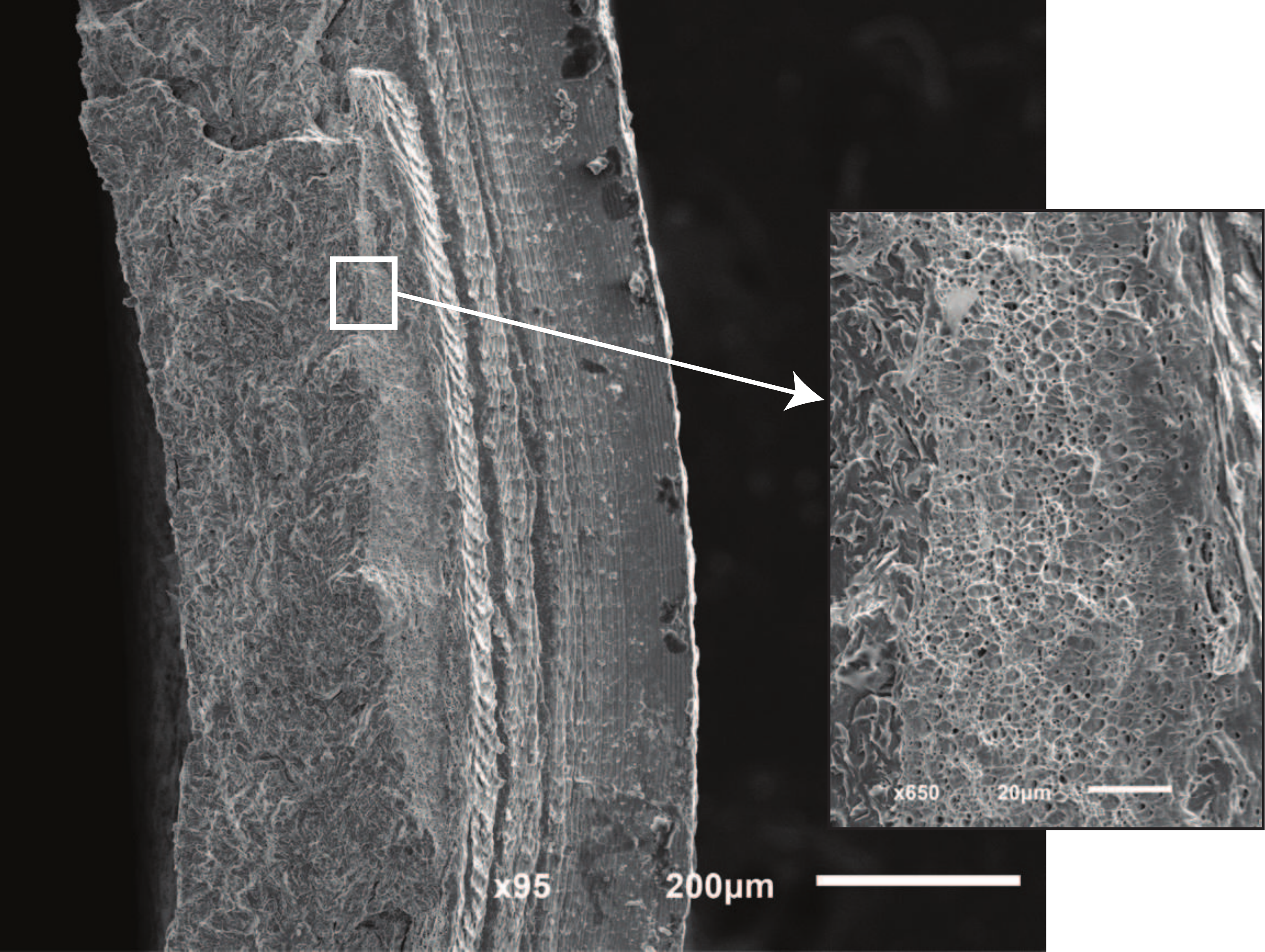}
\caption{SEM image of a CrMoV2 interrupted SPT specimen ($d=0.47$ mm) after breakage inside liquid nitrogen}
\label{fig:WeldVoids}
\end{figure}

The divergence is particularly significant in the case of the weld metal (CrMoV2) so SPT specimens from interrupted tests are split inside liquid nitrogen with the aim of gaining insight of the fracture micromechanisms developed. Fig. \ref{fig:WeldVoids} shows a SEM image of a fractured SPT specimen that has been interrupted at a punch displacement of $d=0.47$ mm. The figure reveals that - unlike conventional fracture tests (see Fig. \ref{fig:FracCrMoV2}) - microvoids are dominant in small punch experiments. Hence, the higher differences observed between standard and SPT critical measurements are justified as, in addition to the aforementioned stress triaxiality disparity, different fracture micromechanisms develop.

\section{Discussion}

Although the present methodology has proven to be able to capture the variation in fracture toughness of different grades of CrMoV steels, some uncertainties remain. 

\begin{figure}[H]
\centering
\includegraphics[scale=0.8]{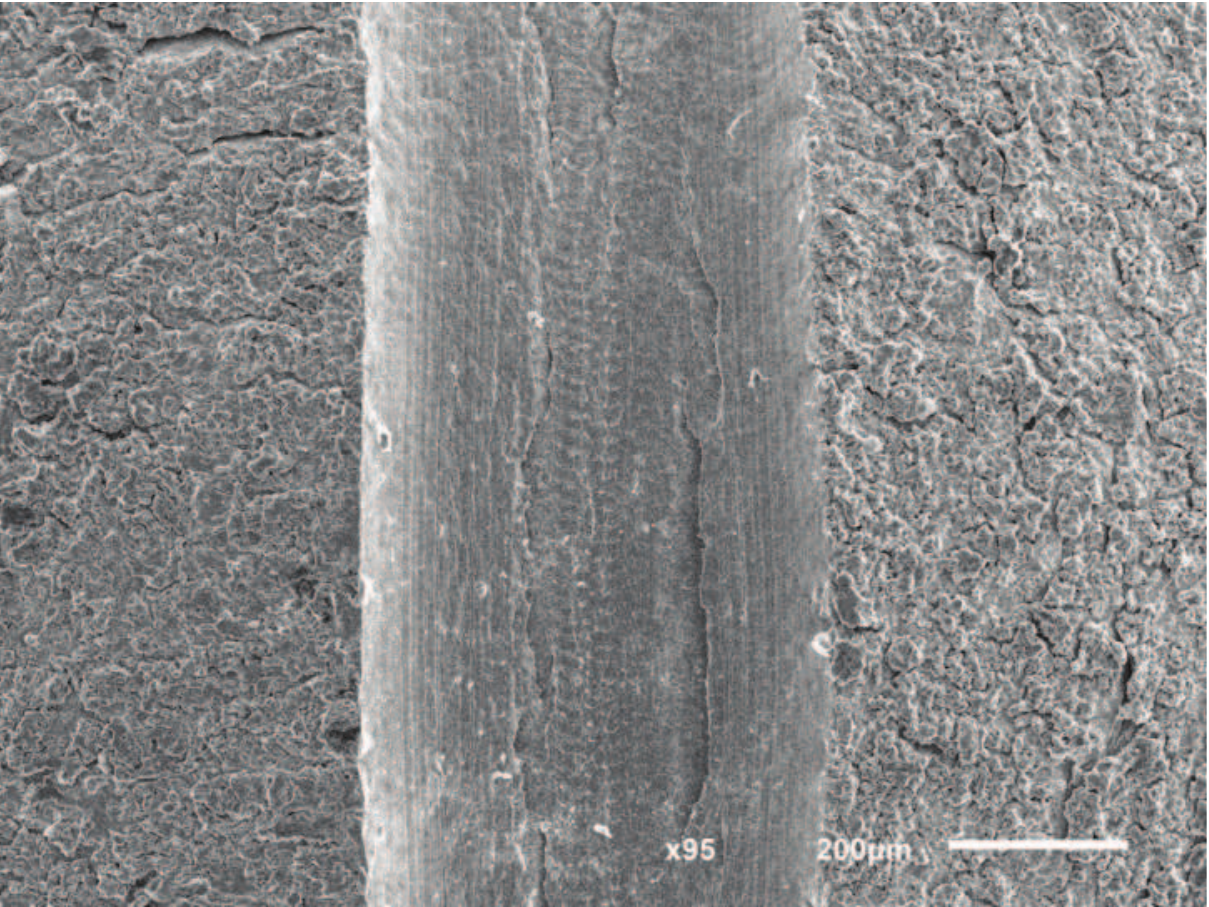}
\caption{SEM image of the observed micro-cracks in CrMoV2}
\label{fig:Microcracks}
\end{figure}

First and foremost, the complex stress state inherent to the SPT does not necessarily favor the formation of a single crack. Moreover, although progress has been made in its micromachining, the notch tip radius is still significantly blunted as compared to a fatigue precrack. Thus, while the final crack always starts at the tip of the notch \cite{G15b}, several microcracks may develop in the vicinity. This latter feature may be particularly relevant in metals with lower fracture toughness, which exhibit brittle behavior in conventional tests. This is the case of the weld metal examined, where - unlike the base metal - the interrupted experiments reveal the existence of microcracks in the early stages of loading (see Fig. \ref{fig:Microcracks}). Such earlier cracking phenomena is not observed in the numerical model and therefore the information provided by a purely microvoid-based damage model must be used with care. 

Capturing the (main-crack driven) final breakage is ensured by fitting the experimental load-displacement curve to back-calculate the damage parameters. However, estimating a critical void volume fraction from void coalescence ahead of the main crack may hinder the modelization of other cracking mechanisms. Thus, the $\delta^{\mathrm{SPT}}_c$ estimated by the numerical model under such circumstances may not be directly comparable to experimental measurements where brittle crack growth occurs. Moreover, as a consequence of the specimen size, metallurgical variability or size effects may play a relevant role in the modelization of damage \cite{MB15}.

\section{Conclusions}
\label{Concluding remarks}

A combined numerical/experimental framework for fracture toughness assessment in notched small punch test (SPT) specimens is developed. With the aim of easing the correlation with conventional fracture tests, the notch mouth displacement $\delta^{\mathrm{SPT}}$ is measured by means of interrupted tests and subsequently compared to the conventional crack tip opening displacement (CTOD). Furthermore, by employing a ductile damage model a critical value of $\delta^{\mathrm{SPT}}$ can be estimated from the onset of void coalescence ahead of the main crack.\\

The present methodology has proven to be able to classify two different grades of CrMoV steel as a function of its fracture resistance. Direct comparisons with standard test measurements are hindered by the different stress triaxiality and damage mechanisms involved. Employing the proposed procedure to examine more metallic materials will very likely contribute to the development of an appropriate scheme for fracture toughness characterization within the SPT.
 
\section{Acknowledgments}
\label{Acknowledge of funding}

The authors gratefully acknowledge financial support from the Ministry of Science and Innovation of Spain through grant MAT2011-28796-CO3-03. E. Mart\'{\i}nez-Pa\~neda also acknowledges financial support from the University of Oviedo through grant UNOV-13-PF. T.E. Garc\'{\i}a additionally acknowledges financial support from the Principado de Asturias Regional Government through the Severo Ochoa Scholarship Programme (BP12-160)




\end{document}